\documentclass[preprint,proceedings]{rmaa}

\usepackage{paralist}

\usepackage{psfrag,color}

\SetYear{2003}
\SetConfTitle{Compact Binaries in the Galaxy and Beyond}

\title{Exploring the Nature of Weak Chandra Sources near the Galactic Centre} 

\author{
  R. M. Bandyopadhyay,\altaffilmark{1} 
  K. M. Blundell,\altaffilmark{1}
  Ph. Podsiadlowski,\altaffilmark{1}
  J. C. A. Miller-Jones,\altaffilmark{1}
  Q. D. Wang,\altaffilmark{2}
  W. N. Brandt,\altaffilmark{3}
  S. Rappaport,\altaffilmark{4}
  and E. Pfahl\altaffilmark{5}}

\altaffiltext{1}{University of Oxford, UK.}
\altaffiltext{2}{University of Massachusetts, Amherst, MA, USA.}
\altaffiltext{3}{Pennsylvania State University, University Park, PA, USA.}
\altaffiltext{4}{MIT, Cambridge, MA, USA.} 
\altaffiltext{5}{Harvard CfA, Cambridge, MA, USA.} 

\shortauthor{Bandyopadhyay et al.}
\shorttitle{Weak Chandra Sources near the Galactic Centre}

\fulladdresses{
\item R. M. Bandyopadhyay, K. M. Blundell, J. C. A. Miller-Jones, and 
  Ph. Podsiadlowski Dept. of Astrophysics,
  University of Oxford, Keble Road, Oxford OX1 3RH, UK
  (\email{rmb, kmb, jcamj, podsi@astro.ox.ac.uk}).
\item W. N. Brandt: Dept. of Astronomy \& Astrophysics, The Pennsylvania
  State University, 525 Davey Lab, University Park, PA 16802, USA 
  (\email{nbrandt@astro.psu.edu}).
\item E. Pfahl: Harvard-Smithsonian Center for Astrophysics, 60 Garden St.,
  Cambridge, MA 02138, USA (\email{epfahl@cfa.harvard.edu}).
\item S. Rappaport: Center for Space Research, Massachusetts Institute of
  Technology, 77 Massachusetts Ave., Cambridge, MA 02139, USA 
  (\email{sar@space.mit.edu}).
\item Q. D. Wang: Dept. of Astronomy, University of Massachusetts,
  Amherst, MA 01003, USA (\email{wqd@astro.umass.edu}).
}

\listofauthors{R. M. Bandyopadhyay, K. M. Blundell, Ph. Podsiadlowski, J. C. A. Miller-Jones, Q. D. Wang, W. N. Brandt, S. Rappaport, \& E. Pfahl}
\indexauthor{Bandyopadhyay, R. M.}
\indexauthor{Blundell, K. M.}
\indexauthor{Podsiadlowski, Ph.}
\indexauthor{Miller-Jones, J. C. A.}
\indexauthor{Wang, Q. D.}
\indexauthor{Brandt, W. N.}
\indexauthor{Rappaport, S.}
\indexauthor{Pfahl, E.}

\abstract{We present early results from the first IR imaging of the
weak X-ray sources discovered in a recent {\it Chandra} survey towards
the Galactic Centre.  From our VLT observations we will identify
likely counterparts to a sample of the hardest sources in order to
place constraints on the nature of this previously unknown
population.}

\addkeyword{Stars: binaries}
\addkeyword{Stars: infrared}
\addkeyword{Stars: Mass loss}
\addkeyword{Stars: X-rays}

\begin{document}
\maketitle

\section{{\it Chandra} Galactic Centre Survey}
\label{sec:intro}

An imaging survey with {\it Chandra}/ACIS-I of the central
0.8$\times$2$^{\circ}$ of the Galactic Centre (GC) revealed $\sim$800
previously undiscovered discrete sources with X-ray luminosities of
$10^{32}-10^{35}$ ergs s$^{-1}$ (Wang et al. 2002).  Our calculations
suggest that the extragalactic contribution to the hard point source
population over the entire survey is $\leq$10\%.  The harder ($\geq$3
keV) X-ray sources (for which the softer X-rays have been absorbed by
the ISM) are likely to be at the distance of the GC, while the softer
sources are likely to be foreground X-ray active stars or cataclysmic
variables (CVs) within a few kpc of the Sun.  These hard, weak X-ray
sources in the GC are therefore most likely a population of X-ray
binaries (XRBs); candidate classes include quiescent black hole (BH)
or neutron star low-mass XRBs, CVs, and high-mass wind-accreting
neutron star binaries (WNSs).

\begin{figure*}[!t]
  \includegraphics[width=\textwidth]{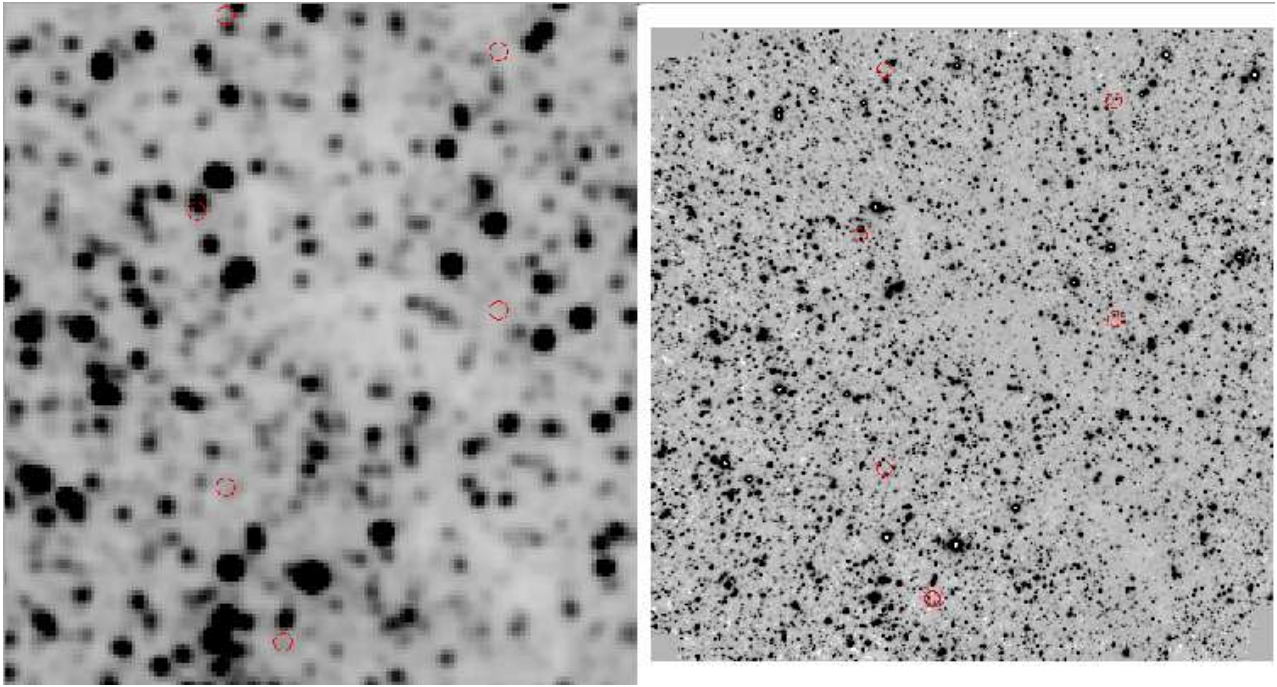}
  \caption{2MASS (left) and VLT (right) $K$-band images of one of our
GC fields.  The circles ({\it Chandra} error circles with 2" radius)
indicate the positions of X-ray sources.  The brightest stars are
clearly visible in both the 2MASS and VLT images; however, the VLT
image is considerably superior in depth and resolution.}
  \label{fig:fig1}
\end{figure*}

\section{What are these Point Sources?}
\label{sec:identity}

Pfahl et al. (2002) have considered the likely nature of these sources
and concluded on the basis of binary population synthesis models that
many of these systems are WNSs.  Depending on the mass of the
companions, the WNSs may belong to the ``missing'' population of
wind-accreting Be/X-ray transients in quiescence or the progenitors of
intermediate-mass XRBs (IMXBs; donor mass 3-7{\mbox{$M_{\odot}$}}).
This {\it Chandra} survey may contain as many as 10\% of the entire
Galactic population of WNSs.  In addition to the WNSs, Pfahl et
al. estimate that a small fraction of these sources could be CVs or
transient BH binaries.

\section{Our VLT Imaging Program}
\label{sec:imaging}

To identify counterparts to a statistically significant number of the
X-ray sources, using ISAAC on the VLT we obtained high-resolution {\it
JHK} images of 26 fields within the {\it Chandra} survey region,
containing a total of 70 X-ray sources.  For the early-type donors of
the WNSs, we expect intrinsic magnitudes of $K$=11-16; these are
therefore readily distinguishable from the majority of late-type
donors expected for BH X-ray transients which have $K\geq$16 in
quiescence.  The average extinction towards the GC is $K\sim$3;
therefore by imaging to a limit of $K$=20 we should detect most of the
WNSs.

The successful achievement of our goals requires astrometric accuracy
and high angular resolution to overcome the confusion limit of the
crowded GC.  There is no archival systematic IR imaging survey which
we could use for this project.  The 2MASS survey has a limiting
magnitude of $K$=14.3, and although the astrometric positions are
accurate to 0.2'', the survey has a spatial resolution of $\geq$2".
As such, the 2MASS data are severely confusion limited in the GC and
moreover are of insufficient depth to detect the majority of the
expected counterparts.  Our VLT images are not confusion limited
(Figure 1) and are complete to $K\sim$20.  For 90\% of the X-ray
sources in our VLT fields, there are one or more resolved IR sources
within each {\it Chandra} error circle.

The distribution of X-ray colours suggests that only a small fraction
of the {\it Chandra} sources are foreground objects.  The next steps
in the analysis of these datasets are (1) to derive accurate
astrometric solutions for both the IR and X-ray images; (2) to
determine IR colours for the potential counterparts within the X-ray
error circles; and ultimately (3) to identify candidates for IR
spectroscopic follow-up observations in order to establish accurate
spectral types for this population.

\end{document}